\documentclass[11pt, a4paper, twocolumn]{article}
\usepackage{caption}
\usepackage{subcaption}
\usepackage{comment}

\usepackage{graphicx}
\usepackage{amssymb}
\usepackage[margin=2cm]{geometry}
\usepackage{authblk}
\title{Predicting nuclear masses with product-unit networks}
\author[1]{Babette Dellen}
\author[1]{Uwe Jaekel}
\author[2]{Paulo S.~A. Freitas}
\author[2,3]{John W. Clark}
\affil[1]{University of Applied Sciences Koblenz, 53424 Remagen, Germany}
\affil[2]{University of Madeira, Campus Universitário da Penteada, 9020-105 Funchal, Portugal}
\affil[3]{Washington University in St. Louis, St. Louis, MO 63130, USA}
\date{}              

\begin{document}
\maketitle

\begin{abstract}
Accurate estimation of nuclear masses and their prediction beyond the 
experimentally explored domains of the nuclear landscape are crucial to an 
understanding of the fundamental origin of nuclear properties and to many 
applications of nuclear science, most notably in quantifying the $r$-process 
of stellar nucleosynthesis. Neural networks have been applied with some 
success to the prediction of nuclear masses, but they are known to have 
shortcomings in application to extrapolation tasks. In this work, we propose 
and explore a novel type of neural network for mass prediction in which the 
usual neuron-like processing units are replaced by complex-valued product 
units that permit multiplicative couplings of inputs to be learned from the 
input data. This generalized network model is tested on both interpolation 
and extrapolation data sets drawn from the Atomic Mass Evaluation. Its 
performance is compared with that of several neural-network 
architectures, substantiating its suitability for nuclear mass prediction.  
Additionally, a prediction-uncertainty measure for such complex-valued 
networks is proposed that serves to identify regions of expected low 
prediction error.
\end{abstract}

\section{Introduction}
\label{sec:intro}
The mass of a nucleus $(Z,N)$, specified by its respective proton and neutron 
contents, is arguably its most fundamental experimentally derived property, 
because the difference from the sum of its proton and neutron rest masses 
determines its binding energy due to interactions between these constituents. 
Often this difference is referred to as the mass defect.  Its accurate 
measurement and theoretical prediction are central to a deep understanding 
of the origin of the elements as well as fundamental nuclear interactions, 
and ultimately important for many applications of nuclear science 
\cite{Mumpower_2016,Kawano}.

With respect to theoretical prediction, diverse mass formulas as a function 
of $Z$ and $N$ have been proposed or derived, beginning with the 
phenomenological macroscopic liquid-drop model \cite{Gamow,weiz35}, followed 
in time by related semi-empirical macroscopic-microscopic models such as 
the sequences of Finite-Range Droplet Models (FRDM) and Weizs\"acker-Skyrme 
(WS) Models that incorporate specific quantum phenomena such as shell 
stucture and nucleonic pairing.  For information on recent developments 
in this category, see especially 
Refs.~\cite{Wang_2014,Moeller_2016,Liu,ZHANG201438}. 

The third traditional category of mass models rests on fundamental theory
and hence is microscopic, in the sense the models are based on assumed 
nucleon-nucleon interactions.  However, the corresponding Schr\"odinger 
equation for the many-nucleon problem becomes intractable with increasing 
$A=N+Z$ and approximations are necessary. Typical approaches in this category 
are based on density functionals and involve assumed nucleon-nucleon 
interactions that allow mean-field treatments (see, for example, 
Ref.~\cite{Gorielyetal2016}).  True microscopic quantum many-body calculation 
of nuclear masses, notably by quantum Monte Carlo techniques, soon 
becomes intractable with increasing $A$ due to combinatoric explosion.
Developments in quantum computation show promise of overcoming  
this barrier \cite{DOEreport}.

An important goal of mass models is to achieve and an accuracy level of
predictions below a root mean squared (RMS) error of $0.100$~MeV, in order to be useful for research on the 
r-process, which is responsible for the creation of about half of the
nuclei heavier than iron.  Among the traditional methods indicated above, 
the best performance in this respect may be represented by the WS4 model 
in the Weizs\"aeker-Skyrme series, still well short of the target. 

Machine learning (ML) techniques provide a quite different approach to the 
prediction of nuclear properties that has gained wide acceptance in the last 
decade and seen rapid growth of applications in diverse variants 
\cite{RMPAIinNP}.  Such methods have the advantage that, beyond the raw 
data, only minimal additional knowledge of the relevant laws of physics 
is required for their implementation. Multilayer neural networks have 
served as a prime example, with successful applications to nuclear masses 
and other properties beginning in the early 1990s 
\cite{Gazula92,Gernoth93,Athan04,Athan06}.  For recent applications of 
neural networks and related ML techniques for prediction of nuclear masses, 
see especially 
Refs.~\cite{Niu,PhysRevC.106.014305,Li2022DeepLA,Neufcourt,Neufcourt2}.   

Feedforward neural networks with at least one hidden layer have been of 
particular interest because they have been proven to be universal function 
approximators \cite{cybenco1989} that can adapt to any kind of data. However, 
a successful predictive model may involve a large number of adjustable 
parameters that must be estimated by optimization. A recently proposed 
neural network for mass prediction achieved RMS errors below $0.4$~MeV on 
training data, but required the fitting of close to a million model 
parameters \cite{Li2022DeepLA}. In contrast, 
Lovell et al.~\cite{PhysRevC.106.014305} developed a probabilistic neural 
network for mass prediction that incorporated theoretical knowledge by 
providing components of phenomenological models as input features. With 
this strategy the number of neurons in the network (and hence the number 
of parameters) could be reduced significantly, while achieving mass 
prediction with RMS errors in the range $0.56-3.9$~MeV. (It must be noted, 
however, that RMS errors reported for different mass studies
depend crucially on the choices of data sets and input features and 
accordingly must be interpreted and compared with caution.)

Beyond describing available data with high accuracy using the previously 
mentioned approaches, there is a critical need for reliable extrapolation 
to lesser known regions of the nuclear chart.  While a particular model may 
perform very well on the data set that was used for determining the model 
parameters, its performance is, in general, likely to fall off beyond this 
domain. For example, traditional neural networks have been shown to perform 
poorly in extrapolation tasks \cite{haley1992,dellenetal2019}. In part, this 
can be understood as a consequence of thresholding: Thresholds and 
threshold-like activation functions used in neural networks compartmentalize 
the feature space and provide, in some simplification, a piece-wise linear 
approximation of the input-output relationship which impairs the extrapolation 
capabilities of the network.

\begin{figure}
\centering
\includegraphics[width=0.8\columnwidth]{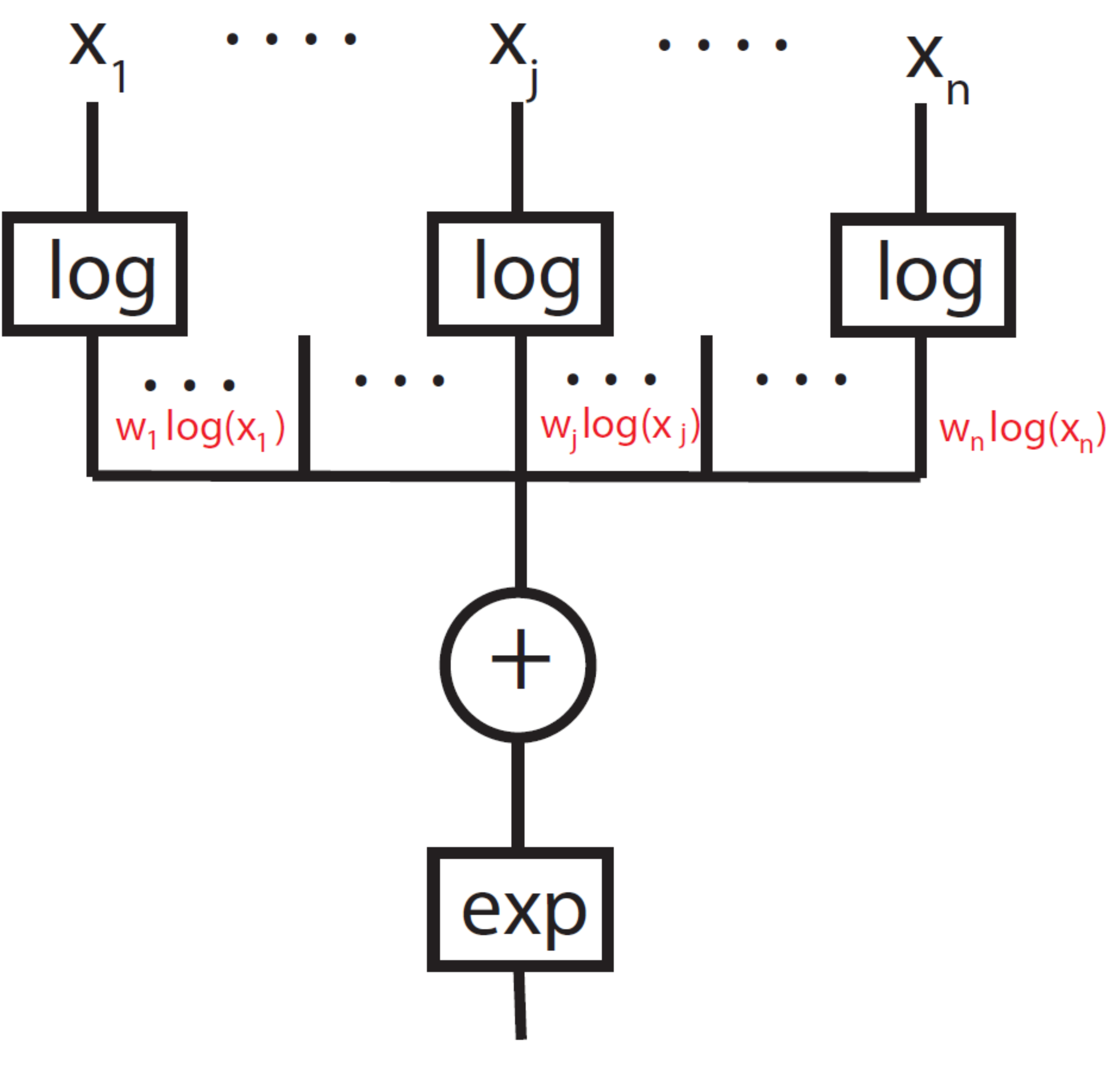}
\caption{Product unit using the logarithmic and 
exponential function as activations to implement multiplicative couplings \cite{dellenetal2019}. In the complex case, $w_j\in \mathbb{C}$ and $x_j\in \mathbb{C}\setminus \{0\}$ for $j=1,...,n$.}
\label{fig1}
\end{figure}

A recent study indicates that product-unit networks  are superior to standard 
feedforward neural networks in function and pattern extrapolation tasks 
\cite{dellenetal2019}. Such networks are composed of nonlinear processing 
units that enable representation of multiplicative couplings of inputs 
\cite{durbin,leerink,dellenetal2019} (see Fig.~\ref{fig1}). Since the information capacity 
of each nonlinear unit is expected to be larger than that of ordinary linear 
units, product-unit networks require fewer neuronal units than classical 
neural networks in modeling the same function. Moreover, the trained weights 
of the product units provide useful information about correlations between 
input features, potentially furnishing insight into the physical laws 
underlying the measured data and contributing to model discovery.  

These special advantages motivated us to adapt and further develop product-unit
networks for the prediction of nuclear masses and to evaluate their performance
in both interpolation and extrapolation tasks.  In doing so, we were led quite 
naturally to propose a novel complex-valued product-unit network that 
does not require thresholds or threshold-like activations, thereby further 
improving performance in extrapolation, their elimination being an advantage in 
itself as indicated above. 
Therefore we compare the proposed complex-valued product-unit network to several alternative real-valued product-unit networks operating with thresholds and classical multilayer perceptrons. 
A further novelty of this generalization is
that the introduction of a prediction-uncertainty measure derived from the 
imaginary spill of the complex-valued network output allows one to identify 
low-error predictions beyond the experimentally explored domain. 

\section{Methods}
\subsection{Real-valued product-unit networks}
Product units compute the products of inputs $x_j$, where $j=1,...,n$, raised 
to the power of their weights $w_j$, yielding $\prod_j x_j^{w_j}$. They can in fact be implemented with conventional neural networks by using both logarithmic and exponential functions as 
activations, as shown in Fig.~\ref{fig1} \cite{dellenetal2019}. For example, the product $a^2b$ can be represented as $\exp{(2\log{a} + \log{b}})$. By 
combining product units in parallel and summing their weighted outputs, 
input-output relationships of the type $\sum_k \alpha_k \prod_j x_j^{w_{jk}}$ 
can be described \cite{dellenetal2019}. Such networks can be trained using gradient descent and 
other standard optimization methods.  A disadvantage of real-valued product 
units is that they are not defined for non-positive inputs.  However, if the 
input is directly passed to a layer composed of product units, this constraint 
can be addressed by appropriately shifting the input values.  If real-valued 
product units are being incorporated into larger network structures, 
thresholds have to be applied that allow only positive data to pass.  
\begin{table*}
\caption{Specifications of various networks built from product units (PU), 
linear units (LU) and ReLU activations (ReLU). CPUN-XS, CPUN-S and CPUN-L are complex-valued product unit networks of different sizes.}
\begin{center}
\begin{tabular}{|c|c|c|c|c|c|c|c|c|}
\hline
Layer & RPUN & CPUN-XS & CPUN-S& CPUN-L &HPUN-S& HPUN-L& MLP-L &MLP-XL \\
\hline
Input & $2$ & $10$ & $10$& $10$& $10$& $10$& $10$& $10$\\
\hline
Hidden 1 & $50$ PU& $10$ LU& $10$ LU& $20$ LU  & $10$ LU & $20$ LU& $20$ LU& $40$ LU \\
 & & & &  & + ReLU& + ReLU & + ReLU&  + ReLU\\
\hline
Hidden 2 & x & $10$ PU& $20$ PU & $40$ PU & $20$ PU& $40$ PU& $40$ LU & $40$ LU \\
 & & & &  &  &  & + ReLU & + ReLU\\
\hline
Hidden 3 & x & x & x& x& x& x& x& $40$ LU\\
 & & & & &  &  &  & + ReLU\\
\hline
Output & $1$ LU & $1$ LU& $1$ LU& $1$ LU& $1$ LU& $1$ LU& $1$ LU& $1$ LU\\
\hline
\hline
\end{tabular}
\label{tab1}
\end{center}
\end{table*}

The architectures of various real-valued networks used in this work are 
summarized in Table~\ref{tab1}.  The network RPUN consists of $50$ product 
units followed by a single linear output unit. No thresholds are applied. 
This network is defined only for positive inputs. The networks HPUN-S and 
HPUN-L are hybrid networks that are composed of a linear layer with 
ReLU-activation, i.e., $f(x) = \max{(\epsilon,x)}$, where $\epsilon$ is a very small number, followed by a product-unit layer and a single linear output 
neuron. This network structure allows transformations of the input data to 
be learned as well. The ReLU-activation ensures that only positive outputs 
are passed to the next layer.

For comparative purposes during evaluation, we also investigated two classical 
multilayer perceptrons with ReLU activation. The network MLP-L contains two 
hidden layers and the network MLP-XL has three.  

To optimize the real-valued networks, we use squared differences as 
loss function. For the RPUN network (implemented in Matlab) optimization was 
performed using stochastic gradient descent and training until convergence for 
$20,000$~epochs. For the hybrid networks and the multilayer perceptron 
(implemented in Pytorch), Adam-optimization (including a learning-rate 
scheduler) with a batch size of $500$ was used. We trained until convergence 
for $50,000$~epochs. Results may vary pending on initialization and hyperparameter choices. 

\subsection{Complex-valued product-unit networks}
Another and indeed more elegant way to handle non-positive data is to 
introduce complex-valued product-unit networks, i.e., networks with 
complex-valued weights and complex activations. To enable linear transformations of the input data, a linear complex-valued preprocessing layer is added to the network. In complex space, the 
logarithmic function is also defined for negative inputs. However, the 
non-zero constraint still applies. Interestingly, this rarely poses a 
problem, because in complex space, weights do not have to cross the origin 
to change the sign of their real part during optimization.  Application of 
complex-valued product-unit networks thus allows one to incorporate 
processing units without having to include thresholds. Here we investigate 
complex-valued product-unit networks of three different sizes, named CPUN-L, 
CPUN-S, and CPUN-XS (see Table~\ref{tab1}). Their output is a complex number 
$z = x+iy$. To optimize the network, we therefore use $(x-\hat{x})^2 + y^2$ 
as loss function, where $\hat{x}$ is the real-valued experimentally measured 
mass defect of the nucleus. These networks are implemented in Pytorch with 
Adam-optimization using a batch size of $500$ and a learning-rate scheduler. 
The real part of the output is used to represent nuclear mass. Results may vary pending on initialization and hyperparameter choices. 

In experiment~B, we trained for an additional $10,000$~epochs to average the 
RMS errors and the prediction uncertainties over this additional training 
period.  In doing so, small random effects induced by the batch size which 
are more likely to occur in product-unit networks could be removed.

\subsection{Data sets}
The atomic mass evaluation AME2020 provides atomic mass values together with 
their measurement uncertainties \cite{Huang_2021,Wang_2021}.  This data 
collection was used to generate a training data set Train-A composed of $2827$ 
nuclei and an interpolation test set Inter-A composed of randomly selected 
$730$ nuclei (roughly $25\%$). These data were used in \textit{Experiment~A}. 

In \textit{Experiment~B}, we explored the extrapolation capabilities of 
the networks of Table~1. Here we used, in addition, 
an earlier data set, namely FRDM2012 \cite{Moeller_2016}, in providing 
experimentally established nuclear masses. From this resource, $2149$ data 
points were split into a training set train-B containing $2099$ data points 
and an interpolation test set Inter-B composed of $50$ randomly selected 
nuclei. Also selected is an extrapolation test set Extra-B composed of 
nuclei of the AME2020 data set that have a mass uncertainty value smaller 
than $0.5$~MeV and are not part of Train-B. The test set Extra-B contains 
$1022$ data points.  So as not to impair extrapolation performance by 
reducing the size of the training set, Inter-B contains only a small amount 
of data. 

For comparative evaluation, we generated in Experiment~C another pair of 
training and test sets by splitting the data set Train-B randomly into two 
sets.  Train-C contains $1949$ nuclei and Inter-C, $200$.  Similarly, $336$ 
nuclei were selected from AME2020 to generate the extrapolation test set 
Extra-C1.  Only nuclei with a mass uncertainty smaller than $0.17$~MeV entered 
the latter data set, while limiting the range of $N$ and $Z$ values. This 
was done to obtain data sets similar to those used in \cite{Li2022DeepLA}.  
We also generated a second extrapolation test set to include nuclei with 
larger $N$ and $Z$, resulting in the set Extra-C2 containing $44$ nuclei.  
The color-coded map of the data sets involved in Experiment~C is shown in 
Fig.~\ref{fig:ExpC}d.

\section{Results}
\subsection{Experiment~A: Input features and multiplicative couplings}
\begin{table}
\caption{Experiment~A: Standard deviation (RMS error) in MeV (rounded) for the 
data sets Train-A and Inter-A, as obtained with the RPUN and CPUN-L networks.}
\begin{center}
\begin{tabular}{|c|c|c|}
\hline
Network & $\sigma_{TrainA}$ & $\sigma_{InterA}$ \\
\hline
RPUN & $2.86$ & $2.87$ \\
CPUN-L & $0.51$ & $0.78$\\
\hline
\end{tabular}
\label{tab2}
\end{center}
\end{table}

\begin{figure}
\centering
\includegraphics[width=\columnwidth]{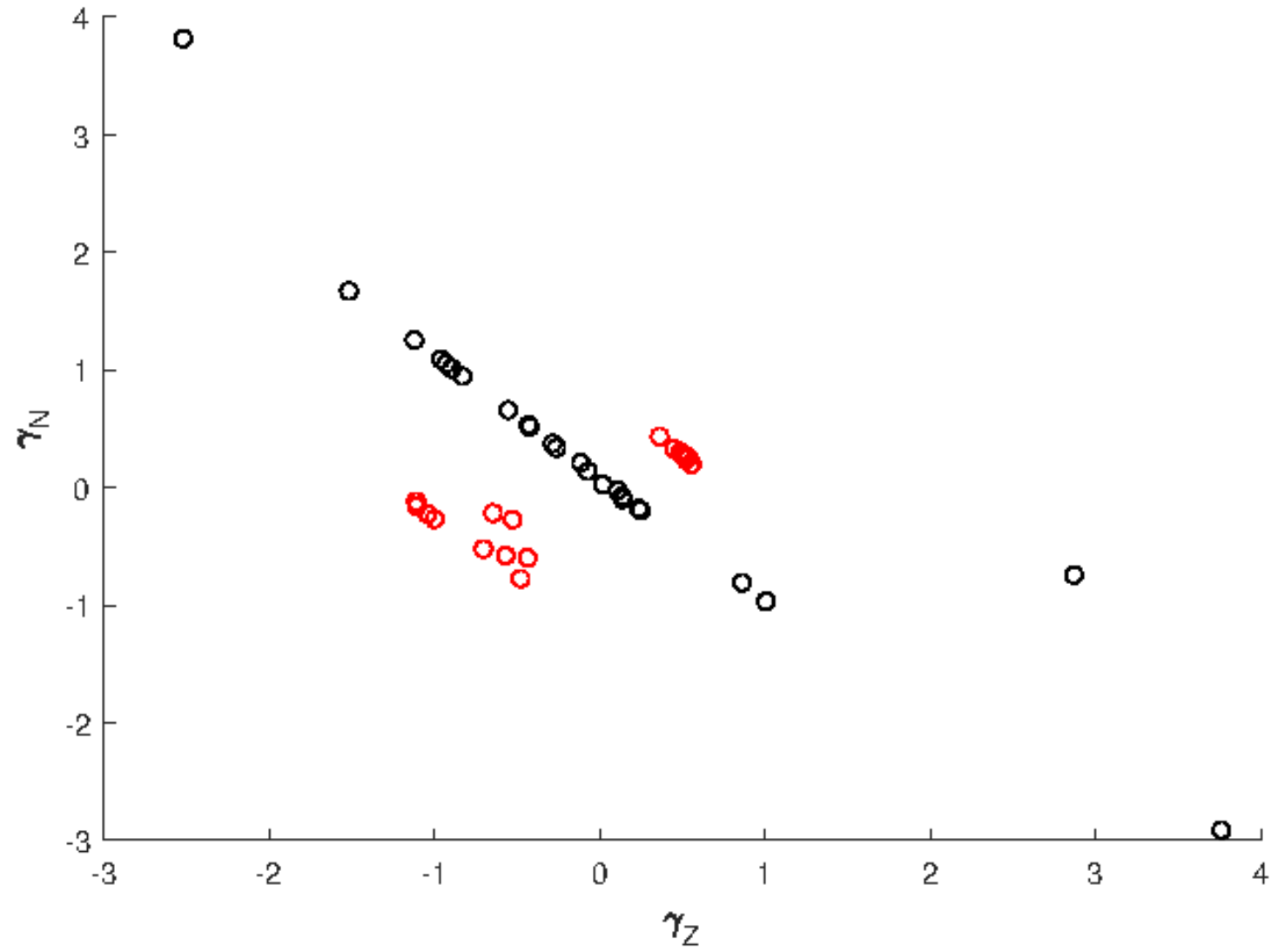}
\caption{Experiment~A: Exponents of the product units of the real-valued 
product-unit network RPUN are shown in exponent space. The network RPUN 
revealed in some approximation a functional dependence of the mass excess on 
$Z$ and $N$ of the form $f(Z/N)g(N)$, where f and g are functions.}
\label{fig:exponents}
\end{figure}
Here we first investigated a real-valued product-unit network of $50$ product 
units that has as input variables only the proton and neutron numbers 
$Z$ and $N$ (network RPUN), a very basic exemplar.  Using stochastic gradient 
descent, convergence was achieved after $20,000$ epochs. The standard deviation 
was computed for the training and test sets, i.e., Train-A and Inter-A, as 
shown in Table~\ref{tab2}. The standard deviations (RMS errors) for the 
training and interpolation test data are in the same range, indicating 
little or no overfitting.  Even though the RMS error is still large, this 
network outperforms the probabilistic neural network M2 of 
Ref.~\cite{PhysRevC.106.014305}, in which an RMS error of $3.9$~MeV was 
reported for a training set of $2074$ nuclei from the 2016 Atomic Mass 
Evaluation (AME2016). However, these results cannot be directly compared, 
since we use a larger training set of $2827$ nuclei that also contains mass 
values afflicted with large uncertainty. 

\begin{figure*}
\centering
\includegraphics[width=\textwidth]{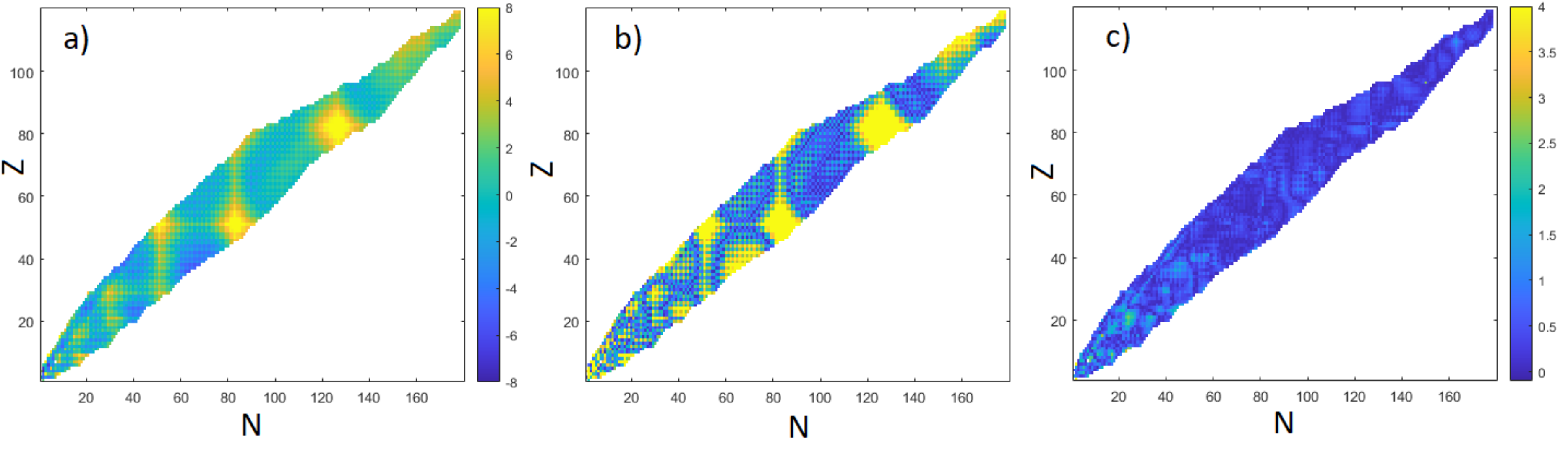}
\caption{Experiment~A: (a-b) Differences between predicted mass and true mass 
in MeV show larger errors in the vicinity of magic numbers. The finer patterns 
in the differences repeat with respect to parity. (c) Including additional 
features (magic numbers, parity) and using the CPUN-L model improves 
predictions. Larger differences are observed for small atomic numbers.}
\label{fig:ExpA}
\end{figure*}

In Fig.~\ref{fig:exponents}, the exponents of the product units are shown 
for $Z$ and $N$. Exponents of product units with a negative weight factor are 
shown as red circles; those with positive ones as black circles. The size of 
the circles increases with increasing weight factor. Some circles are 
coincident due to product units having the same exponents.  A regular 
structure is revealed that suggests a functional dependency of the mass 
excess on $Z$ and $N$ of the form $f(Z/N)g(N)$, where $f$ and $g$ are 
functions. The heat map for the differences between predicted and true 
mass values as a function of $Z$ and $N$ (as seen in Fig.~\ref{fig:ExpA}a 
and \ref{fig:ExpA}b) shows large errors in the vicinity of magic numbers. 
Fine repetitive errors indicate a dependence of mass on parity. 

To improve the model, information about the proximity to the magic numbers was 
included with the input data. Accordingly, we added two features to the input 
data, namely the distance to the closest magic number of $Z$ and the distance 
to closest the magic number of $N$, for a number of nuclei. For protons, the 
magic numbers $2,8,20,28,50,82,126,164$ were included, and for neutrons 
neutrons, $2,8,20,28,50,82,126,184,196$.  For each data point, we computed 
first the distance from the closest magic numbers, separately for protons and 
neutrons. Additionally, we included features 
indicating odd and even parities for $N$, $Z$, and $A$.  By training the 
network CPUN-L with this extended set of input features, ten in number, 
the predictions could be improved (see Table~\ref{tab2}).  As a result, 
a training RMS error of $0.51$~MeV was achieved.

The absolute mass difference is shown in Fig.~\ref{fig:ExpA}c. Compared to 
the model RPUN, the error in the vicinity of the magic numbers has decreased,
as has that associated with parity effects.  Larger errors above $1$~MeV 
are more frequently observed for small atomic numbers. This may indicate that 
the input features selected are not carrying sufficient information to model 
the data more accurately in that region. 

\subsection{Experiment~B: Mass extrapolation and prediction uncertainty}
\begin{table}
\caption{Experiment~B: RMS error in MeV (rounded) for the CPUN-L network for 
different thresholds $\tau$.}
\begin{center}
\begin{tabular}{|c|c|c|c|}
\hline
Data set & $\tau$ & $N$ & $\sigma$\\
\hline
Extra-B &- & $1022$ & $4.12$ \\
Extra-B &$5$ & $959$ & $1.85$\\
Extra-B &$0.1$ & $256$ & $0.55$ \\
Extra-B &$0.04$ & $30$ & $0.39$ \\
\hline
Inter-B &- & $50$ & $0.36$\\
Inter-B &$0.1$ & $44$ & $0.18$\\
\hline
Train-B &$-$ & $2099$ & $0.15$\\
\hline
\end{tabular}
\label{tab3}
\end{center}
\end{table}

We have investigated the extrapolation capability of the complex-valued 
product-unit network CPUN-L, trained with the data set Train-B, containing 
$2099$ nuclei. Unlike the data set used in Experiment~A, the training 
data now includes only nuclei with well-established mass values of low 
uncertainty.  After 60,000 epochs, an RMS error of $0.15$ is achieved for 
the training set (see Table~\ref{tab3}), outperforming recent approaches 
evaluated on similar data \cite{PhysRevC.106.014305,Li2022DeepLA}. 
In particular, Ref.~\cite{Li2022DeepLA} reported a training error of 
$0.263$~MeV for a similar data set by predicting the residual between 
the liquid-drop-model prediction and the true mass, using a deep neural 
network containing over a million parameters.  In contrast,
an RMS error of $0.56$~MeV has been achieved far more economically
with a probabilistic neural network~\cite{PhysRevC.106.014305}. 

\begin{figure}
  \begin{subfigure}[b]{\columnwidth}
    \centering
    \includegraphics[width= \columnwidth]{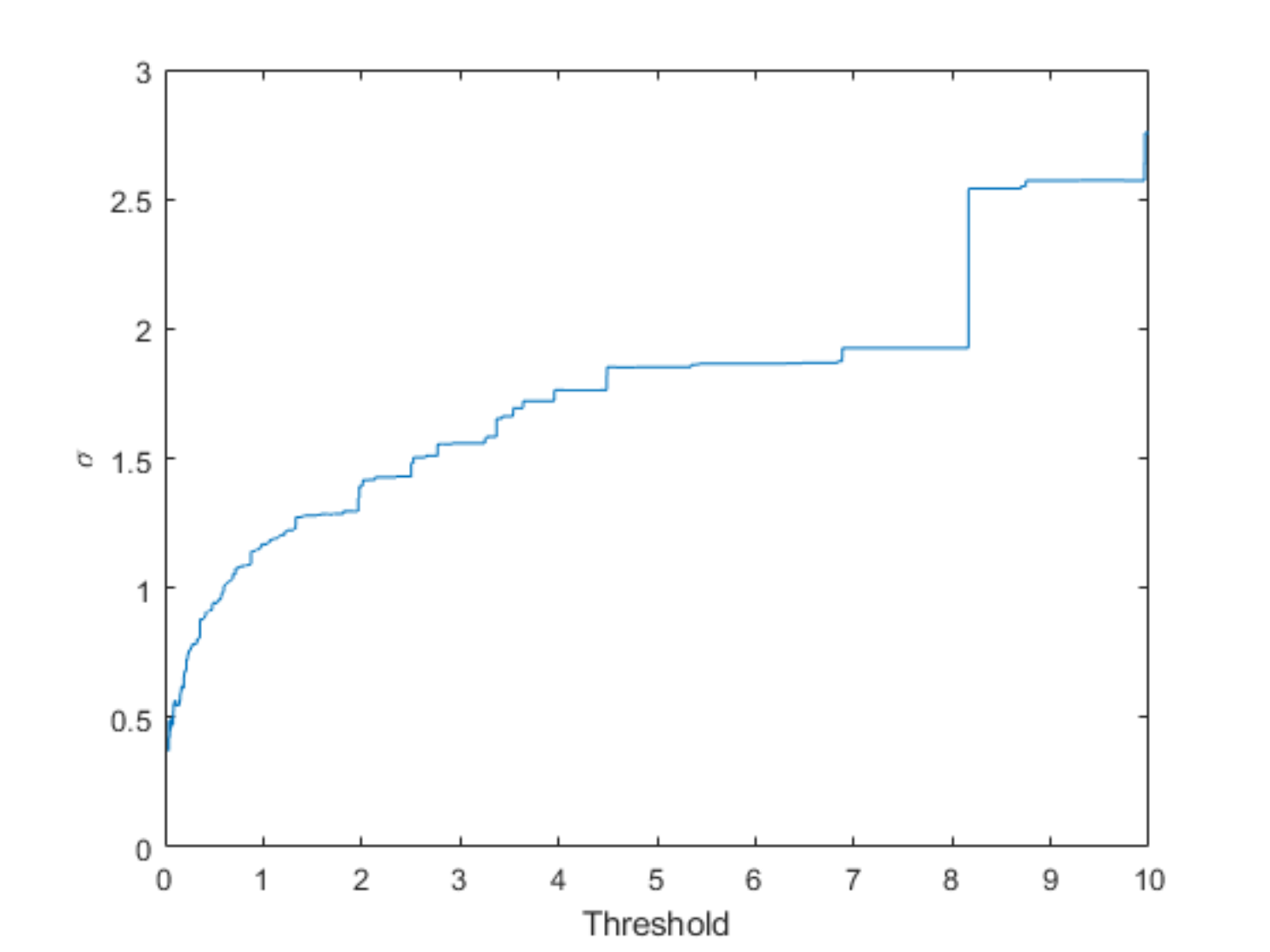}
    \caption{RMS error vs threshold}
    \label{fig:extra_error}
  \end{subfigure}
  \hfill
  \begin{subfigure}[b]{\columnwidth}
    \centering
    \includegraphics[width = \columnwidth]{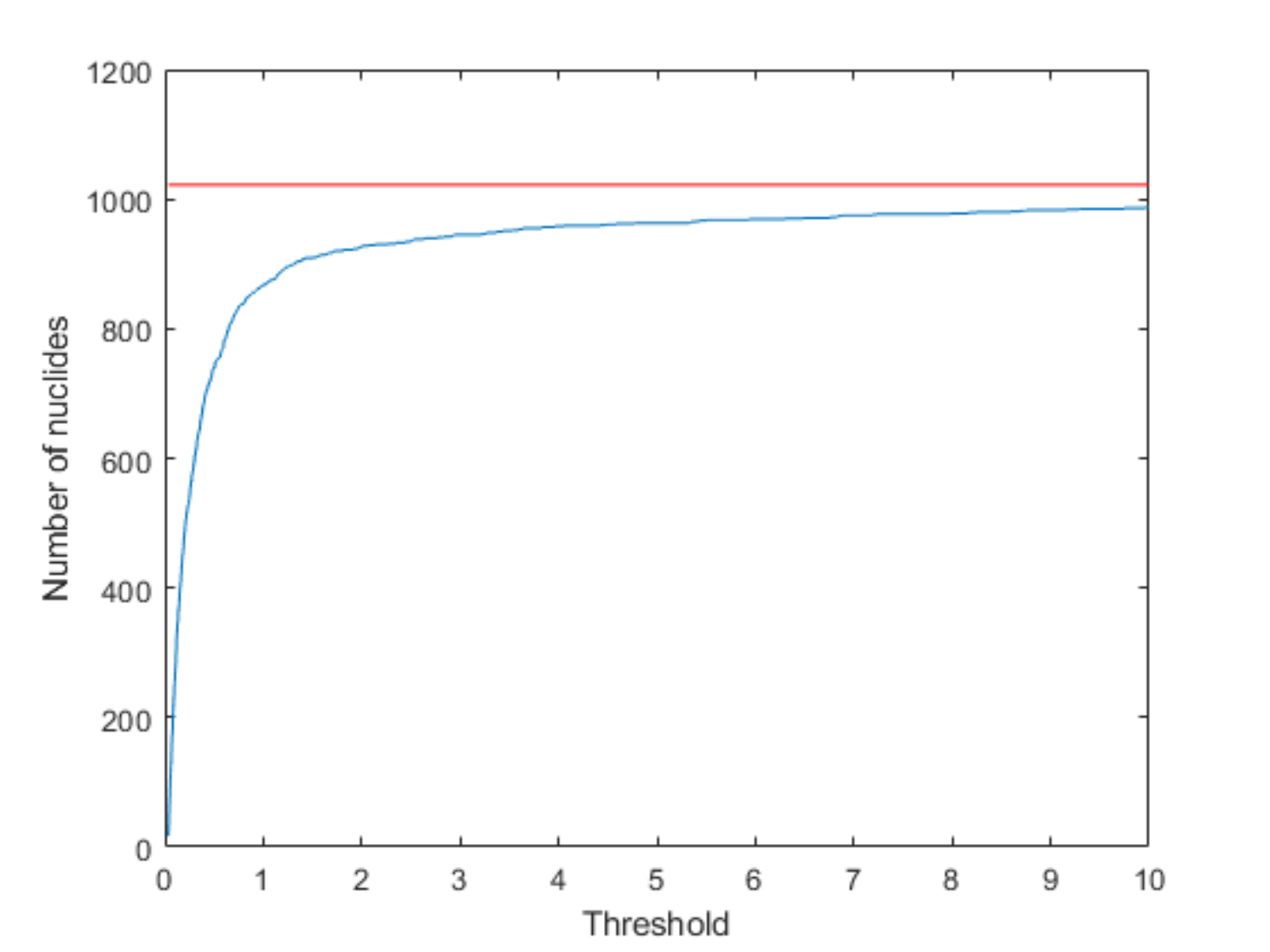}
    \caption{Number of nuclides vs threshold}
    \label{fig:extra_np}
  \end{subfigure}
\caption{Experiment~B: (a) RMS error in MeV plotted as a function of 
threshold $\tau$. (b) Size of data sets in number of nuclei, plotted as a 
function of $\tau$.}
\label{fig:net_sig5}
\end{figure}

Since performance evaluation on training data has limited significance for
performance on other data, we have evaluated model performance further on a 
large extrapolation test set containing $1022$ nuclei partially afflicted 
with large uncertainty, but below $0.5$~MeV.  We also evaluated interpolation 
performance using a small interpolation test set containing $50$ nuclei. 
As expected, the RMS error is larger for the test data.  An RMS error of 
$4.12$~MeV was obtained for the large extrapolation data set, and an RMS error 
of $0.36$~MeV for the interpolation test set. 

\begin{figure*}
\centering
\includegraphics[width=\textwidth]{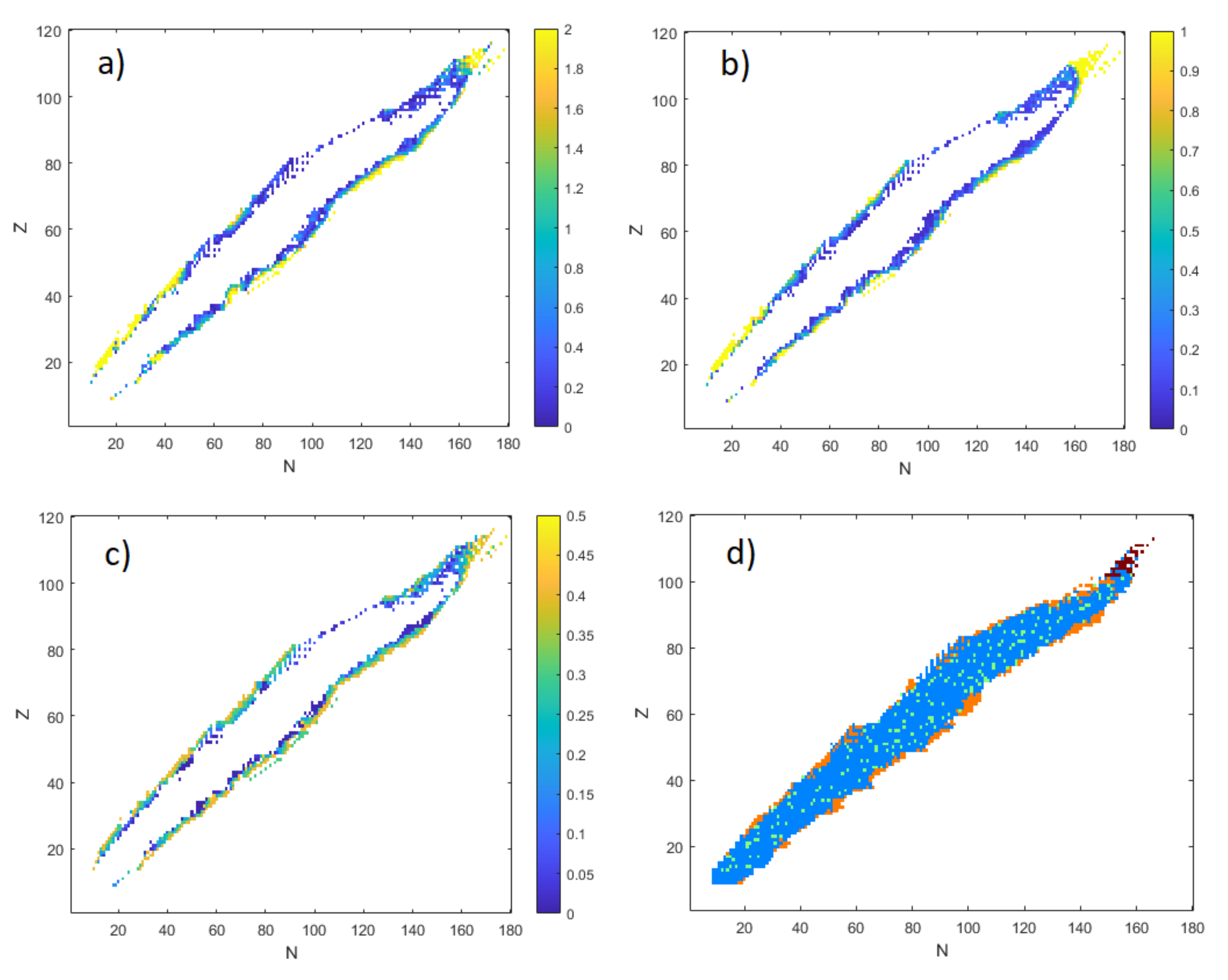}
\caption{(a) Absolute difference between predicted and true masses in MeV 
in Experiment~B.  (b) Absolute value of the imaginary part of the output 
of CPUN-L provides a prediction uncertainty for predicted masses in 
Experiment~B.  (c) Heat map of measurement uncertainties (AME2020) in MeV 
for the masses in the extrapolation region. (d) Training and test sets used 
in Experiment~C: Train-C (blue), Inter-C (light blue), Extra-C1 (orange), 
Extra-C2 (dark red). 
}
\label{fig:ExpC}
\end{figure*}

Importantly, however, we can identify subsets of test data with a lower 
RMS error by using the imaginary component of the output to measure prediction 
uncertainty. In Figure~\ref{fig:ExpC}a, the absolute mass differences between 
predicted and true values are shown for the extrapolation set.    
Fig.~\ref{fig:ExpC}b shows the absolute value of the imaginary component 
of the output, defining the prediction uncertainty. It provides a useful 
indicator for erroneous predictions. 

By applying a threshold $\tau$ to the prediction uncertainty, we can select 
subsets of nuclei and calculate the RMS error for a chosen subset. In 
Figure~\ref{fig:extra_error} the RMS error is plotted as a function of 
$\tau$, and seen to decrease with $\tau$. The size of the subsets decreases 
as well, as shown in Figure~\ref{fig:extra_np}. A similar behavior is 
observed for the interpolation set (not shown). The results for a set of 
thresholds are summarized in Table~\ref{tab3}. For the interpolation test 
set, it is possible to predict masses that have not been included in the 
training set with an RMS error below $0.2$~MeV. For extrapolation, the 
RMS error for mass prediction could be reduced to $0.39$~MeV. 

\begin{table*}
\caption{Experiment~C: RMS error in MeV (rounded) for the training and test 
sets, for different networks.}
\begin{center}
\begin{tabular}{|c|c|c|c|c|c|c|c|}
\hline
Data & CPUN-XS & CPUN-S& CPUN-L &HPUN-S& HPUN-L& MLP-L &MLP-XL \\
\hline
Train-C & $0.54$ & $0.37$ & $\mathbf{0.17}$& $0.56$& $0.38$& $ 0.77$& $0.37$ \\
\hline
Inter-C & $0.49$& $0.38$ & $\mathbf{0.26}$  & $0.60$ & $0.45$& $0.90 $ & $0.51$ \\
\hline
Extra-C1& $0.73$& $0.66$ & $\mathbf{0.39}$& $1.55$ &$0.98$ & $1.58$ &$1.23$  \\
\hline
Extra-C2& $1.23$ & $0.74$& $\mathbf{0.65}$& $2.72 $& $1.13$& $1.79$& $3.53$ \\
\hline
\end{tabular}
\label{tab4}
\end{center}
\end{table*}

\subsection{Experiment~C: Performance evaluation and comparison}
This experimental sequence is devoted to further comparison of the 
performance of the network CPUN-L and the other network architectures 
specified in Table~\ref{tab1}.  The extrapolation data set used for 
Experiment~B contains many nuclei measured with a large mass uncertainty 
(see Fig.~\ref{fig:ExpC}c). By keeping only nuclei with mass uncertainty 
smaller than $0.17$~MeV, an extrapolation data set presumably similar to that used 
in Ref.~\cite{Li2022DeepLA} could be assembled (see orange-colored nuclei 
in Figure~\ref{fig:ExpC}d). We also evaluated relative performance for 
an additional extrapolation set containing nuclei with large atomic 
masses (nuclei colored dark red). The interpolation test data is shown 
in light blue, the training data in darker blue. 

The results after $50,000$ epochs are shown in Table~\ref{tab4} for the 
different networks listed.  The network CPUN-L outperforms the other networks 
on all data sets, achieving an RMS error of $0.17$~MeV for the training set, 
$0.39$~MeV for the extrapolation test set Extra-C1, and $0.65$~MeV for the 
extrapolation test set Extra-C2. The deep multilayer perceptron MLP-XL 
achieved a training and interpolation error in the range of most product-unit 
networks (except CPUN-L), but is less suited to predict nuclear masses at 
larger distance from the training set than the product-unit networks. We 
suspect that the use of thresholds (ReLU activations) impairs the 
extrapolation capabilities of the MLP.  

The network CPUN-L also outperforms a recent deep-learning model 
\cite{Li2022DeepLA} in the extrapolation task. Ref.~\cite{Li2022DeepLA} 
obtained an RMS error of $0.263$~MeV for the training set and $0.605$~MeV 
for an extrapolation set that we presume to be approximately identical with 
data set Extra-C1.

\section{Conclusion}
We have proposed a novel type of neural network for nuclear-mass prediction. 
The model is composed of product units that allow multiplicative couplings 
of inputs to be learned from the data. This mechanism enables the capture of 
hidden structures that are not accessible by classical neural networks. 
By working with complex-valued weights and activations, we could propose a 
product-unit network (CPUN-L) that delivers predictions with an RMS 
error well below $0.2$~MeV in the experimentally explored domain of the 
nuclear chart, as well as for a group of selected nuclei within a random 
test set. We have further identified groups of nuclei in the extrapolation 
region with an RMS error below $0.4$~MeV. Regions of low-error predictions 
were identified using a novel prediction-uncertainty measure that is specific 
for this type of network. 

We have compared a variety of network structures and shown that the 
complex-valued network CPUN-L performs best in terms of training RMS error 
and both interpolation and extrapolation test-set RMS errors. The networks used 
in this study have been small compared with deep networks currently employed
for similar tasks \cite{Li2022DeepLA}. It is certainly possible to further 
improve training performance by adding more layers and neurons (thus 
parameters) to the networks studied here, including the classical multilayer 
perceptron, but only at the risk of serious overfitting, i.e., an increase 
of the RMS error for the test data, especially the extrapolation test set.

\end{document}